%% file: amdahl.tex
\documentclass[preprint]{sig-alternate}
\usepackage{balance}

\usepackage{graphicx,color,subfigure,colordvi,epsfig,textcomp,verbatim,url,
citesort,amsmath,times}

\makeatletter
\def\ps@headings{%
\def\@oddhead{\mbox{}\scriptsize\rightmark \hfil \thepage}%
\def\@evenhead{\scriptsize\thepage \hfil \leftmark\mbox{}}%
\def\@oddfoot{}%
\def\@evenfoot{}}
\makeatother
\newcommand\T{\rule{0pt}{2.6ex}}
\newcommand\B{\rule[-1.2ex]{0pt}{0pt}}

\newlength\SUBSIZE

\begin{document}
\conferenceinfo{DaMoN}{'11 Athens, Greece}
\CopyrightYear{2011} 

\title{Hadoop in Low-Power Processors}

\numberofauthors{3}
\author{
\alignauthor
Da Zheng\\
       \affaddr{Computer Science Department}\\
       \affaddr{Johns Hopkins University}
       \affaddr{Baltimore, MD 21218, USA}\\
       \email{dzheng5@jhu.edu}
\alignauthor
Alexander Szalay\\
       \affaddr{Department of Physics and Astronomy}\\
       \affaddr{Johns Hopkins University}
       \affaddr{Baltimore, MD 21218, USA}\\
       \email{szalay@jhu.edu}
\alignauthor Andreas Terzis\\
       \affaddr{Computer Science Department}\\
       \affaddr{Johns Hopkins University}
       \affaddr{Baltimore, MD 21218, USA}\\
       \email{terzis@jhu.edu}
}
\date{25 March 2011}

\maketitle

\input{abstract}




\input{intro}

\input{twoapp}

\input{exp}

\input{revisit}

\input{concl}


\bibliographystyle{abbrv}
\bibliography{amdahl}

\balancecolumns

\end{document}

%% file: abstract.tex
\begin{abstract} 
In our previous work we introduced a so-called {\em Amdahl blade}
microserver that combines a low-power Atom processor, with a GPU and
an SSD to provide a balanced and energy-efficient system. Our
preliminary results suggested that the sequential I/O of Amdahl blades
can be ten times higher than that a cluster of conventional servers
with comparable power consumption. In this paper we investigate the
performance and energy efficiency of Amdahl blades running Hadoop. Our
results show that Amdahl blades are 7.7 times and 3.4 times as
energy-efficient as the Open Cloud Consortium cluster for a
data-intensive and a compute-intensive application, respectively. 
The Hadoop Distributed Filesystem has relatively poor performance on
Amdahl blades because both disk and network I/O are CPU-heavy
operations on Atom processors. We demonstrate three effective
techniques to reduce CPU consumption and improve performance. However,
even with these
improvements, the Atom processor is still the system's bottleneck. We
revisit Amdahl's law, and estimate that Amdahl blades need four Atom
cores to be well balanced for Hadoop tasks. 
\end{abstract}

%% file: intro.tex
\section{Introduction}

The volume of data that scientific instruments generate is doubling
every year~\cite{Bell09}. In turn, the need to process this constantly
increasing amount of data at the same or even higher rates has lead to an
unsustainable increase in the power consumption of compute clusters
for data-intensive applications.

In an attempt to tackle this power consumption issue, Szalay et
al. recently introduced the {\em Amdahl blade}
concept~\cite{blade}. This microserver combines an energy-efficient
CPU (e.g., Intel Atom) with a GPU and a SSD to build a {\em balanced},
in terms of processing and I/O rates, and energy-efficient
system. Preliminary results from the same work suggested that a
cluster of Amdahl blades can be up to ten time more efficient than
existing Beowulf clusters with the same I/O rates for a set of
sequential and random disk access patterns~\cite{blade}. On the other
hand, Reddi et al. showed that while Atom processors are more
energy-efficient than Xeon processors for tasks such as web searches,
the overall system is less efficient because of platform
overheads~\cite{webatom}. 

This paper evaluates the performance and energy efficiency of the
Amdahl blades when running Hadoop~\cite{hadoop}, the popular
open-source implementation of MapReduce, for scientific
applications. To do so, we implemented a data-intensive and a
compute-intensive astronomy application and compare the
performance of a cluster of Amdahl blades to an Open Cloud Consortium
(OCC) cluster.

The experimental results show that the Amdahl blades are approximately
7.7 times as energy-efficient as the OCC cluster for
the data-intensive application and 3.4 times as efficient for the
compute-intensive application. Moreover, the experiments show that
Amdahl blades are CPU-bounded. The reason is that disk and network I/O
operations are surprisingly CPU-heavy on Atom processors. In this
sense the performance of the whole system can be improved by using
more powerful Atom processors. We estimate that a quad-core Atom
processor should be enough to build a balanced Amdahl blade for
Hadoop.

We also find that the performance of the Hadoop Distributed Filesystem
(HDFS) is vital to data-intensive applications, but it has poor
performance on the Amdahl cluster, due to the limitations mentioned
above. The paper demonstrates some effective methods to improve the
performance of HDFS.  Specifically, reducing the overhead of the Java Native Interface can
improve the performance of the data-intensive application by up to a
factor of two, while LZO compression and direct IO can improve its
performance by 61\% and 37\%, respectively, when the replication
factor is 3. The observation that compression can improve performance
might be surprising, when the system is CPU-bounded. However,
considering that both disk and network I/O consume considerable CPU
time, compression can reduce overall CPU consumption by reducing the
amount of data written to the disk and the network. 

Shafer et al. also proposed mechanisms for improving the performance
of Hadoop Distributed Filesystem~\cite{hdfsrice}. Their methods,
however, focus on improving the disk performance, so they might not
improve the performance of our system.  We, on the other hand,
investigate the impact of Atom processors on HDFS, and try to improve
its performance by reducing CPU consumption.  

%


\label{sec:intro}

%% file: twoapp.tex
\section{Two applications}
\label{sec:twoapp}

We use two  astronomy applications to measure the energy
efficiency of the Amdahl blades. To make the comparison with the OCC 
cluster more comprehensive, the first application is
data-intensive while the second is compute-intensive. 

\subsection{Neighbor Searching}
\label{sec:neighborsearch}

The first application reports all the neighbors of each object on the sky in an
astronomy dataset that are within a user-defined radius $\theta$. All
objects in the dataset are on the surface of a sphere. We
re-implemented the Zones algorithm \cite{zones}, which was originally 
implemented in SQL.


The MapReduce implementation divides the surface of the sphere
into blocks of equal size. The task of the mappers is then to partition
the data and copy it in a way that guarantees that each reducer has a
complete block of data. Specifically, mappers assign each object in
the original dataset a block ID and move all objects with the same ID
together. To simplify searching for neighbors of objects that are
close to each block's borders, the mappers also copy objects that are
within a certain region around each block. 

Each reduce invocation processes all objects in a block and the
objects from its neighboring blocks and outputs all object pairs
within a certain distance.  Intuitively, the amount of computation
increases with the block size. On the other hand, as the size of each
block decreases the total number of blocks increases and thus the
amount of border block data that need to be copied increase.  An
optimization that we employ is to have the reducer process larger
blocks and split each block further before calculating the distance of
all object pair; first, the reducer calculates the distance between
every two objects in the same sub-block and then between objects in a
sub-block and objects in its neighboring sub-blocks. This optimization
is very effective when $\theta$ is small and the implementation always
favors larger blocks. In this case, the application becomes less
compute-intensive. Instead, when objects on the sphere become very
dense, the output size becomes very large, and the application becomes
data-intensive.  For example, the size of the current input dataset is
approximately 25GB, and the application outputs 540GB data when
$\theta = 60''$. We note that the application still involves
considerable computation, and it can become more compute-intensive as
$\theta$ decreases.

\subsection{Neighbor Statistics}

The second application uses the same input data and is similar to the
first one. The difference is that instead of outputting all object
pairs within a certain distance, it computes the distribution of the
number of object pairs in terms of distance. For example, our
implementation calculates the number of pairs for $\theta \in
\{1'',2'',3'',\ldots,60''\}$

This application includes two MapReduce steps. The first step uses the
same customized input function and the same map function as the 
previous application and the reducer uses the same algorithm to
compute the distance using the previously described
optimization. However, each reducer only outputs the statistics for
each block. Since the amount of output data is very small, reducers
produce text output for simplicity. The second MapReduce step is very
simple: mappers parse the data from the previous step and a single
reducer combines all data and outputs aggregated statistics. This
application is very compute-intensive.

%% file: exp.tex
\section{Evaluation}
\label{sec:exp}

Next, we measure the performance of the two applications on a cluster
of Amdahl blades and on the OCC cluster,
located at University of California, San Diego. Before measuring the 
performance of two applications, we first measure the disk performance
of a single Amdahl blade and the performance of the Hadoop Distributed
Filesystem (HDFS).  We also tune Hadoop, mainly HDFS, to optimize its
performance for data-intensive applications and use the same
configuration on the OCC cluster. 

\subsection{Amdahl Blade Configuration}

\begin{figure*}[htbp]
\centering
\includegraphics[width=\textwidth]{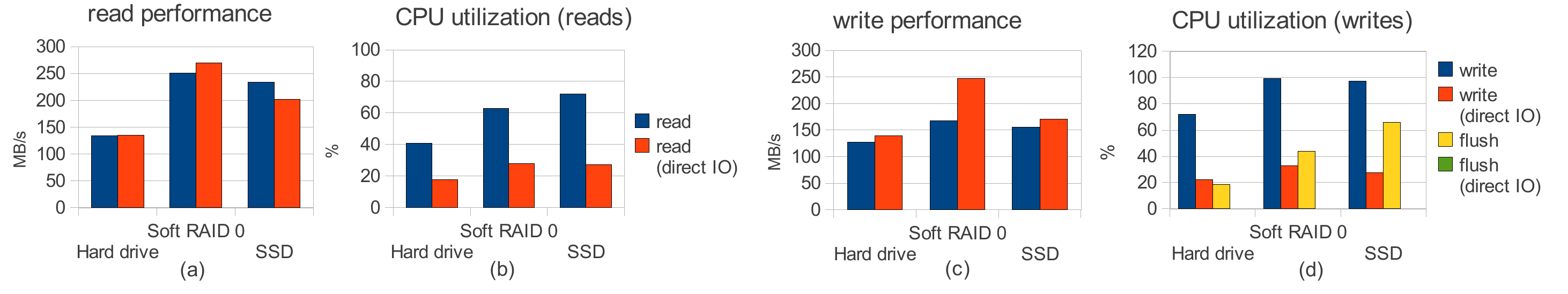}
\caption{Disk I/O performance and CPU utilization. The results
  from Figures~(a) and (c) suggest that direct I/O improves write
  performance, especially for Software RAID 0, while it does not
  improve read performance. Figures~(b) and (d) indicate that direct
  I/O reduces CPU utilization, especially for writing. In Figure~(d),
  0\% of CPU is used for buffer flushes, when direct I/O is used.}
\label{diskio}
\end{figure*}

Each Amdahl blade uses the Zotac IONITX-A platform, containing one 
dual-core Atom 330 processor, clocked at 1.6 GHz, and the Nvidia ION chip
(GeForce 9400M), 4GB memory, two Samsung Spinpoint F1 1TB conventional hard drives
and one OCZ 120GB Vertex drive. Hyperthreading is enabled. A single
48-port 1Gbps Ethernet switch connects all nodes in the
cluster~\cite{blade}. We use a total of nine cluster nodes; one as the
master, and the rest as slaves. All nodes run 64-bit Scientific Linux
6, JVM OpenJDK 1.6, and Hadoop v0.20.2.


It is possible to reduce disk I/O during the data shuffling phase
through proper configuration. After the mappers produce their output,
the data must be sorted and partitioned, before reducers can use
it. Since Hadoop v0.17~(\cite{shuffle}), data shuffling works as
follows: the data that a mapper outputs is held in a memory buffer
with pre-configured size. Hadoop uses two buffers for this
purpose. One buffer stores the output data from mappers, while the
other stores the metadata related to the output data. Whenever the
size of one of the buffers exceeds a threshold, its contents are
sorted and copied to the disk. Once a mapper outputs all of its data,
it performs another mergesort and writes the results to the
disk. If both buffers are large enough, one disk write and one
disk read can be eliminated.

The size of the data that the mapper outputs in our case is slightly
larger than the size of the input, i.e., 64MB. Each input record is 57
bytes and the size for one output record is $57+8=63$ bytes (key +
value). Assuming the number of records increases by 10\% (10\% is a
conservative estimate and the actual number is much smaller), the size
of output data is 77MB.
Hadoop keeps four integers as metadata for a record and therefore the
size of the metadata is 20MB.
The write-to-disk buffer threshold is 80\% by default and therefore
the total buffer size should be at last 125MB.
%
Using these parameters, most mappers only need to write data to the
disk once.  

\begin{table}[tb]
\centering
\begin{tabular}{|l|l|}
\hline
\textbf{parameters} & \textbf{value}\\
\hline
\texttt{{\small dfs.replication}} & {\small 1 or 3}\\
\hline
\texttt{{\small dfs.block.size}} & {\small 64MB}\\
\hline
\texttt{{\small mapred.child.java.opts}} & {\small -Xmx512m}\\
\hline
\texttt{{\small mapred.job.reuse.jvm.num.tasks}} & {\small -1}\\
\hline
\texttt{{\small io.sort.mb}} & {\small 125}\\
\hline
\texttt{{\small io.sort.record.percent}} & {\small 0.2}\\
\hline
\texttt{{\small io.sort.spill.percent}} & {\small 0.8}\\
\hline
\texttt{{\small io.bytes.per.checksum}} & {\small 4096}\\
\hline
\texttt{{\small mapred.tasktracker.reduce.}} & \\
\texttt{{\small tasks.maximum}} & {\small 2 or 3}\\
\hline
\texttt{{\small mapred.tasktracker.map.tasks.maximum}} & {\small 3}\\
\hline
\end{tabular}
\caption{Hadoop configuration parameters.}
\label{config}
\end{table}%

Table~\ref{config} contains all the Hadoop configuration parameters
that we use. In the case of the Neighbor Searching application, each
node runs two reducers because the DataNode process consumes
significant CPU and memory during the reduce phase; for the Neighbor
Statistics application, each node runs three reducers, because very
little data is written to HDFS, and only reducers are active in the
reduce phase.

\subsection{I/O performance on a single Amdahl Blade}

We measure disk performance with a simple Java application which
reads/writes 64 MB of data using a single thread from/to a file for
100 times, each time using a different file, simulating how HDFS reads
data from and writes data to the disk. We collected measurements for
the magnetic disk, SSD, and Linux software RAID 0 on top of the two
magnetic disks on the blade. During writes, data is first copied from
user space to the filesystem cache and from there the kernel's flush
thread writes data to the disk. To capture the relative load of both
operations we measure the CPU usage of the Java program and the flush
thread independently. In addition to normal I/O operations, a Java
program can also use direct I/O to read and write data, using the Java
Native Interface. Since direct I/O requires aligned memory, the
program pre-allocates a piece of aligned memory in the C code and
copies all data between the aligned memory and the Java heap.

As Figure~\ref{diskio} suggests, direct I/O not only improves write
performance, but also reduces CPU use dramatically. Data written to a
file with direct I/O bypasses the filesystem cache and thus the flush
thread is not involved. Considering that reducers write data to HDFS
that is not going to be read  in the immediate future, it makes
sense to use direct I/O. On the other hand, direct I/O provides little
improvement for data reads. 

Direct I/O reduces CPU use for the following reason. During a normal
write, data is copied from the user space to the Linux kernel cache
where it is split into pages. When the number of dirty pages exceeds a
threshold, the kernel starts submitting I/O requests to the disk driver
to write the dirty pages to the disk. Since data was split to pages,
many more disk requests for individual pages are initiated and the
overhead of VFS becomes surprisingly high when running on the Atom
processor. On the other hand, when large blocks of data are written
from the user space with direct I/O, only one write request is sent to
the disk driver thereby avoiding much of the computation overhead.

Network I/O is another kernel operation that consumes much CPU in
Amdahl blades. When a reducer writes data to HDFS, it invokes the HDFS
client interface to send data to the local data node using a TCP
socket. When data is replicated among data nodes in HDFS, data is also
sent using TCP. Finally, reading data from HDFS also involves network
communications.

We used a Java program to measure the network throughput and
corresponding CPU utilization of the Amdahl
blades. Table~\ref{nettable} shows that network I/O, like disk I/O,
generates considerable overhead. Network transmission between
processes on the same node requires three memory copies: from the user
space to the kernel, inside the kernel, and from the kernel to the
user space. In other words, the maximum application rate of 343 MB/s
requires approximately 1 GB/s memory copy rate. The maximal memory
bandwidth measured by SiSoft Sandra is only about 2.6GB/s, which includes
data sent to the cache and data written back to the memory, and the 
maximal memory copy rate we measured is 1.3GB/s; thus, network IO in the 
local case very likely saturates the memory bus. Table~\ref{nettable} also
shows that network transmission to a remote node is more expensive
than local traffic. Unlike disk I/O, there is no way in Linux to
reduce the CPU overhead of network transmission between two nodes. The
only place to reduce the CPU overhead is to let processes at the same
node communicate via shared memory.

\begin{table}[tb]
\centering
\begin{tabular}{|c|c|c|c|}
\hline
{\bf Traffic}& {\bf Max. throughput}& {\bf CPU(send)}&{\bf CPU (receive)}\T\B \\
\hline
{\bf local}  & 343MB/s	& 98.96\%			& 99.27\%\\
{\bf remote} & 112MB/s	& 36.76\%			& 88.1\%\\
\hline
\end{tabular}
\caption{Network I/O is very CPU-heavy on the Amdahl blades.} 
\label{nettable}
\end{table}%

\vspace{10pt}
\subsection{HDFS performance on Amdahl blades}

\begin{figure*}[htbp]
\centering
\includegraphics[width=\textwidth]{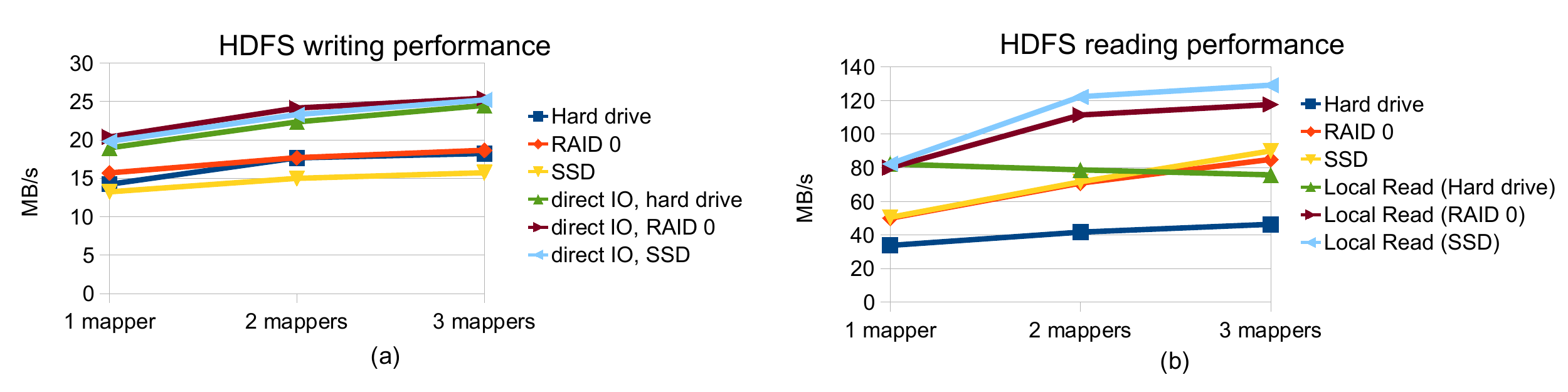}
\caption{HDFS performance on the Amdahl cluster. Figure~(a) shows
  direct I/O is an effective way to improve write performance,
  while different hardware configurations do not affect performance
  significantly. Figure~(b) shows that read performance  from
  the local node is much higher than reading from another
  node. Moreover, HDFS on one hard drive performs significantly worse
  than the other two configurations.}
\label{hdfsfig}
\end{figure*}

Considering the observations from the previous experiments, we expect
that HDFS performance improves with direct I/O and for this reason we
modified HDFS to use it. We measured the performance of HDFS using the
TestDFSIO benchmark, provided with the Hadoop
distribution. Figure~\ref{hdfsfig} presents the measured per-node
throughput when each mapper writes/reads 3GB and the replication
factor is set to three. 

Figure~\ref{hdfsfig}(a) shows that HDFS performs better when using
more than one mapper writing data simultaneously. This result
contrasts the result of Shafer et al. who suggested that concurrent
reads/writes can decrease performance~\cite{hdfsrice}. At the same
time, the performance difference between two and three mappers is
small. The reason is that the system is CPU bounded and more writers
consume more CPU resources. The results also show that while direct
I/O provides considerable benefits, the different hardware configurations
have almost the same I/O performance. Again, the reason is that CPU is
the bottleneck of the system and the only way to improve performance
is to reduce CPU consumption. Even though direct I/O does improve
performance, the throughput of writing to the disk is about 75MB/s,
only half of the throughput of one hard drive. The Java profiler shows
that the DataNode process spends about 80\% of its time on network 
transmission when direct I/O is enabled. In
order to further improve write performance, one should either use a
faster CPU, reduce the size of data transmitted over the network,
or use a different network stack with lower overhead such as TCP-Lite.


While hardware configuration does not affect write performance, it
does have an impact on HDFS read performance. Figure~\ref{hdfsfig}(b)
presents results for two different types of reads: reading from HDFS,
and reading data from HDFS that resides in the same node as the
reader.  Reading from the local node is more relevant to the MapReduce
programming model because the master node of MapReduce always
considers data locality when assigning mapper tasks.  HDFS has much
better performance in reading than in writing, which is not surprising
since HDFS shares the Google file system (GFS) design, and GFS was
designed for append-once-read-many workloads~\cite{gfs}. 

Reading from the local node outperforms reading from other nodes
because reading data from the disk and sending it to the client are
done sequentially in HDFS. Sending data to the client at the same node
is much faster, so more disk I/O requests can be issued within the same time period.  An
interesting observation here is that HDFS on one hard drive has much
worse performance than on the other configurations, and the reason is
that RAID 0 and SSD have much better read performance than a single
hard drive. We can see the performance declining when multiple mappers
read data simultaneously when HDFS runs on a single hard drive. Shafer et al. suggested
that this decrease is due to multiple concurrent readers causing more
disk seeks~\cite{hdfsrice}. The {\em iostat} utility shows that the
hard drives are fully utilized in both cases of one hard drive and RAID 0
when three mappers read data
simultaneously. So the performance can only be improved further by
reducing seek time. One can also improve HDFS performance by
parallelizing disk reads and network transmissions, which can be
achieved with either asynchronous I/O, or using two threads
dedicated to disk reading and network transmission.

Direct I/O is not enabled for reads since, as shown in the
previous section, it does not improve performance
appreciably. More importantly, using direct I/O means that the
application should implement its own prefetching mechanism. Without
prefetch, the problem mentioned above where disk reads and
network operations are done sequentially in HDFS becomes even more
serious. As a matter of fact, our experimental results showed that
direct I/O decreases the reading performance of HDFS significantly.

In summary, HDFS throughput is significantly smaller than that of the
native Linux filesystem. Other than the problems we mentioned above,
HDFS has significant CPU overhead. Two factors contribute the majority of 
the overhead.  First, the Hadoop filesystem is implemented in the
user space and it interacts with other processes via TCP/IP, even for
local processes.  As shown above, network communication has
considerable CPU overhead.  Second, Hadoop generates checksums when
outputting data, and verifies them when receiving data.

\subsection{Improve the Performance of the Neighbor Searching Application}
The Neighbor Searching application, described in
Section~\ref{sec:neighborsearch}, is data-intensive. In this section,
we measure its performance on the Amdahl cluster and discuss methods
that improve HDFS performance, and thus application performance. In 
the experiments of this section and the following ones, we run HDFS 
on Linux software RAID 0, and direct I/O is only enabled for writing.

\subsubsection{Overhead of Java Native Interface}

When data is written to HDFS, one checksum is calculated for a certain
number of bytes (512 bytes by default). The default checksum algorithm
used in Hadoop is CRC32, and it is implemented using the Java Native
Interface (JNI). Whenever a reducer writes bytes to HDFS, it
calculates the checksum. However, it turns out that JNI is very
expensive on the Atom processor. If checksum is calculated each time a
small amount of bytes are written to HDFS, the overhead of calculating
CRC32 will be extremely high.

There a couple few solutions for this: {\em (1)} Use a Java
implementation of CRC32, so the JNI overhead can be avoided.  The
latest version of Hadoop has one such implementation, but we use the
older version of Hadoop because it is considered to be stable. {\em
  (2)} Reduce the number of JNI invocations. Each record output from
the reducers in Neighbor Searching has only 24 bytes, and the original
implementation writes 8 bytes to HDFS each time, which in turn invokes
the JNI function. The number of invocations can be reduced by placing
a new {\tt BufferedOutputStream} on the top of the original {\tt
  OutputStream}. Thus, data is written to HDFS only when the buffer in
{\tt BufferedOutputStream} is full.

As Figure~\ref{figimprove} shows, the second approach improves the
performance of the Neighbor Searching application by a factor of two
when the replication factor is one, and by 47\% when the replication
factor is three. The default HDFS configuration calculates a checksum
for every 512 bytes. The number of bytes can be increased in order to
further reduce the overhead of JNI. Experiments shows the performance
hardly improves further after the number of bytes (specified by
\emph{io.bytes.per.checksum}) reaches 4096.

\subsubsection{LZO}

Hadoop v0.20.2 provides two compression algorithms: Gzip,
Bzip2. However, both of them are CPU intensive and so we use the LZO
algorithm. LZO favors speed over compression ratio, so it is more
lightweight. Nevertheless, it still helps reducing the output size
from the reducers by 60\%.  As Figure~\ref{figimprove} shows, when
the replication factor is one, compression does not improve
performance. However, when the default replication factor is used,
there is significant performance improvement, and the time used by
the reducers decreases to 62\%.

It might be surprising that compression can improve the performance
while the system is CPU-bounded. Considering that both disk IO and
network IO consume much CPU, compression can reduce overall CPU
consumption by reducing the amount of data written to the disk and the
network. Since the Amdahl blades are equipped with a GPU, it would be
better to offload the compression computation to GPU to further
improve performance.

\subsubsection{Direct I/O}
\begin{figure}[tb]
\begin{center}
\includegraphics[scale=0.55]{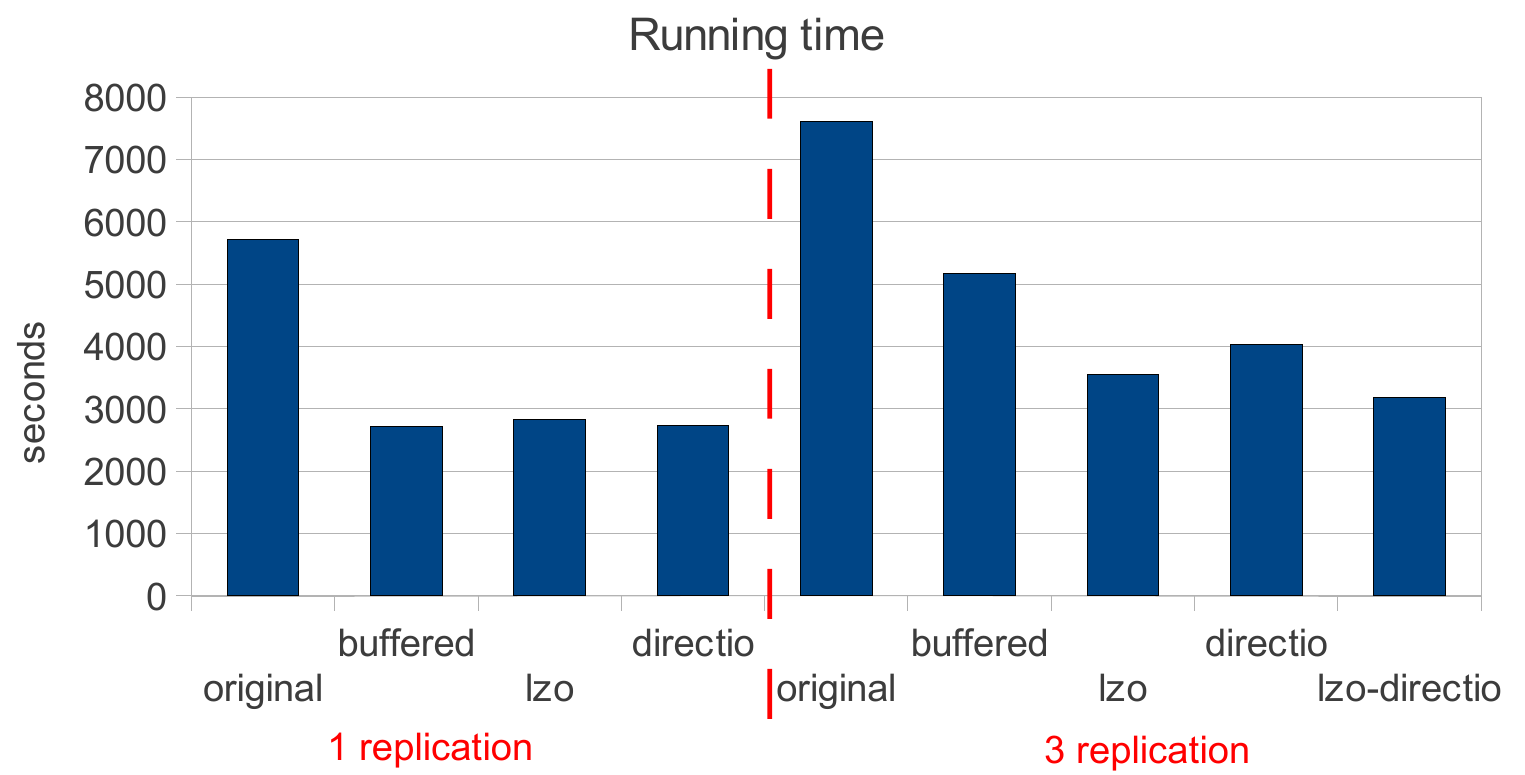}
\caption{Performance improvement of the Neighbor Searching
  algorithm. Buffering output data from reducers to reduce JNI
  overhead improves performance dramatically. Direct I/O and/or
  LZO compression effectively improve performance when the
  replication factor is three.}
\label{figimprove}
\end{center}
\end{figure}

The previous section showed that direct I/O can reduce CPU overhead,
and improve HDFS performance. Figure~\ref{figimprove} illustrates the
impact of direct I/O on the Neighbor Searching application. While
direct I/O cannot improve the performance when the replication factor
is one, it can improve performance by 37\% when the replication factor
is three.

\subsubsection{Discussion}


It is possible to further improve HDFS performance by reducing
network I/O.  When mappers read data, they receive data from the
DataNode process via a socket.  Since most of input data resides in the
local node, network transmission between mappers and the local
DataNode should be avoided. When reducers write data to HDFS, data is
sent via a socket as well, even though reducers and the DataNode
processes are at the same node. Again, the network transmission should
also be avoided by using shared memory. These two improvements will be
our future work.

\subsection{OCC cluster Performance}

Each OCC node is equipped with a dual-core AMD Opteron
Processor 2212, clocked at 2GHz, 12GB RAM and one Hitachi Ultrastar A7K1000 disk. 
The nodes in the local
rack are connected with 1Gbps network, and the link between racks is
10Gbps.

In the following experiments, four nodes in the same rack are used, one as the master
node and the other three as data nodes. The default replication factor
is used, so each node has the same copy of data.


The HDFS read and write throughputs are about 65MB/s and 15MB/s, 
respectively, measured by TestDFSIO, and the throughputs of the local 
disk are about 70MB/s and 50MB/s, respectively. The Hitachi disk has
the transfer rate of 85MB/s at zone 0 (at the edge of the disk), 
and 42MB/s at zone 29 (close to the center of the disk). Considering that
about 80\% of space on the disk has been used and that the filesystem prefers to
first use the zones with better performance, it is not surprising to have the
read and write performance, as we show above. The disks
on the Amdahl blades, on the other hand, are almost empty, so they
have their best performance.

Reducers buffer their data output as on the Amdahl cluster.  Since the
disk is the bottleneck, direct I/O is not enabled in the following
test. We encountered difficulties compiling LZO on the OCC cluster,
and since GZIP and Bzip2 consume too much CPU, compression was
disabled. Since nodes have enough memory and Hyperthreading is
enabled, each node runs three mappers and three reducers. Nodes in the
OCC cluster do not have enough space to store the output of the
Neighbor Searching application when $\theta$ is 60$''$, so the
cluster only runs the application with $\theta =15'', 30''$.

Table~\ref{bothtab} presents the running time in seconds of the
applications on the Amdahl and the OCC clusters.
Since the OCC cluster does not use LZO compression, LZO is not
used in the Amdahl cluster either. Table~\ref{bothtab} suggests that
the Neighbor Searching application runs much faster on the Amdahl
cluster, especially when $\theta$ is large. This is not surprising
since the Amdahl blades are designed for data-intensive
applications. The Amdahl cluster has slightly better performance in
the compute-intensive application, which is a little unexpected, and
suggests Atom processors are very efficient compared to the server
processors.

\begin{table}[tb]
\centering
\begin{tabular}{|c|c|c|c|c|c|c|c|c|}
\hline
{\bf} & {\bf 60''}	& {\bf 30''} & {\bf 15''} & {\bf stat}\T\B\\
\hline
{\bf Amdahl}	& 3933	& 1628	& 1069	& 2157\\
{\bf OCC}	& N/A	& 3901	& 1760	& 2334\\
\hline
\end{tabular}
\caption{The running time in seconds of two applications. Columns  60'', 30''
    and 15'' correspond to the results of Neighbor Searching
    application with $\theta = 60'', 30'', 15''$, respectively. Column
    stat represents the results for the Neighbor Statistics applications.} 
\label{bothtab}
\end{table}%

\subsection{Energy consumption}

Each Amdahl blade consumes $\sim$40W at full load while each node in
the OCC cluster consumes 290W. In other words, one OCC node consumes
the same amount of power as seven Amdahl blades. In terms of energy
efficiency, the Amdahl blades are 7.7 times and 3.4 times as efficient
as the OCC cluster for the data-intensive application (when $\theta$ is $30''$) and the
compute-intensive application, respectively.  The bottleneck of the
OCC cluster is clearly in the disk, so it is not very suitable
for data-intensive applications.


%% file: revisit.tex
\section{Revisiting Amdahl's law}
\label{sec:revisit}

The Amdahl blade experiments showed that the Atom processors used are
not powerful enough to fully utilize even the blade's hard disk. The
maximal read and write throughout, as shown in Figure~\ref{diskio}, is
approximately 300MB/s and 270MB/s, respectively, when software RAID 0
is used (the number doubles when SSD is used in parallel). On the
other hand, the maximal throughout of the disk shown in
Figure~\ref{hdfsfig} is 85MB/s and 75MB/s (the throughout of writing
to the disk is 3 times the throughout of writing to HDFS when the
replication factor is 3).

Therefore, it is reasonable to revisit the Amdahl number, which guided
the design of the Amdahl blades. As stated in Amdahl's law, a balanced
computer system needs one bit of sequential I/O per second per
instruction per second~\cite{law}.  Our previous work consider only
disk I/O~\cite{blade}. However, HDFS employs many network I/O
as well as disk I/O operations, so network I/O should also be included
in the calculation. Table \ref{amdahlnum} shows the Amdahl numbers
with and without network I/O.





The number of instructions per cycle (IPC) per core of the Atom
processor, as shown in Table~\ref{amdahlnum}, is constantly below one,
and IPC of HDFS reading and writing is even lower. The lower IPC of
HDFS read and write operations can be explained by the fact that these
operations involve many memory copies and the CPU is busy with moving
data into and out of cache instead of executing instructions. There
are a few reasons that other cases have low IPC. First, Atom
processors use an in-order architecture, in which cache misses waste
more CPU cycles. Furthermore, Atom processors minimize the use of
specialized execution units in order to reduce power consumption. For
example, the SIMD integer multiplier and Floating Point divider are
used to execute instructions that would normally require a dedicated
scalar integer multiplier and integer divider
respectively~\cite{atom}. Thus, some complex instructions such as
division take many clock cycles to finish. Furthermore, the small
cache of Atom processors leads to more cache misses, which further
hurts IPC~\cite{webatom}.


Table \ref{amdahlnum} shows we should include network I/O in the
Amdahl number calculation. While HDFS reads and writes have an Amdahl
number close to one, when network I/O is not included in the
calculation, the numbers decrease considerably when including network
I/O. It is reasonable that the numbers are lower than one because HDFS
tasks involve many I/O operations. The reducer of the Neighbor
Searching application has an Amdahl number of one, when network
I/O is considered. The Amdahl number of mappers is very high, which
suggests mappers are very compute-intensive. The sources of
computation can be reading data from the local data node via a socket,
verifying checksums, sorting data output from mappers, etc.  The Amdahl
number for the Neighbor Statistics application is irrelevant because
reducers output little data, and the application is deemed to be
extremely compute-intensive.

\begin{table}[tb!]
\begin{tabular}{|l|c|c|c|c|c|}
\hline
{\bf} & {\bf Freq} & {\bf IPC} & {\bf InstrRate} &{\bf AD} & {\bf ADN} \T\B \\
\hline
{\bf HDFS read}		& 0.48	& 0.27	& 421.43	& 1.15	& 0.38 \\
{\bf HDFS write}	& 0.79	& 0.22	& 548.75	& 1.3	& 0.43\\
{\bf Mapper}		& 0.98	& 0.56	& 1751.72	& 12.3	& 6.2\\
{\bf Reducer (stat)}	& 1	& 0.69	& 2196.1	& N/A	& N/A \\
{\bf Reducer (search)}	& 0.98	& 0.48	& 1493.87 	& 2.99	& 1\\

\hline
\end{tabular}
\caption{Amdahl number for different Hadoop tasks.  Freq is
    current CPU frequency/nominal frequency, IPC is instructions per
    cycle at the current frequency per core. InstrRate is the rate
    of instructions executed in the processor (million instructions
    per second). AD is the Amdahl number in terms of disk I/O. ADN is
    the Amdahl number in terms of disk I/O and network I/O.}
\label{amdahlnum}
\end{table}%


Even though it is impossible to achieve a perfectly balanced system,
the Amdahl number can guide node design.  Each node has aggregate disk
I/O of $\sim$300MB/s and a network link of 1Gbps. Furthermore, IPC of
Atom processors is about 0.5 as shown in Table
\ref{amdahlnum}. Considering all these factors, we estimate each node
needs six cores/processors, clocked at 1.6GHz, in order to saturate
both disks and network. However, in Hadoop, we are never able to
saturate disks, while the network link is 1Gbps, because data that
needs to be written to the disk needs to be sent to the network.
Although the ratio of data written to the disk and data sent to the
network varies according to the replication factor of HDFS, we can
assume the disk speed is aligned with the network speed, and estimate
that each node needs four cores/processors.

However, in order to achieve a more balanced system, additional
factors should be considered rather than just CPU speed. For example,
memory speed is another very important factor.  The current system is
very likely to be memory bound for some operations such as HDFS reads
and writes. The maximal memory bandwidth is 2.6GB/s, measured by
SiSoft Sandra, and VTune shows the rate of loading data to cache and
writing to memory gets close to the maximal bandwidth for the case of
HDFS writes. Thus, simply having more CPU cores may not improve the
performance of data-intensive applications, and Amdahl blades need to
be equipped with faster memory and a faster memory bus. On the other
hand, this problem can be alleviated to some extent with software. For
example, passing data between local processes using sockets requires
three memory copies; instead, if local processes communicate through
memory mapping, the number of memory copies as well as CPU utilization
can be reduced significantly. At the same time, it is difficult to
perform some optimizations in Java. For example, we cannot map memory
to the Java heap memory, so memory copy becomes inevitable. Java
itself increases the number of memory operations. For example,
whenever objects, including arrays, are created, the memory containing
the objects is initialized. In summary, memory bandwidth is relevant,
and it should be considered, when building a well balanced computer
system.

While quad-core Atom processors are still hypothetical, we can still use 
existing hardware to achieve a more balanced system. Low-power Xeon
processors already become reality. Xeon E3-1220L processors, announced 
in 2011, have higher CPU frequency than Atom processors and large L3 
CPU cache, and support faster memory. These processors should have 
much higher IPC and thus have much better performance while only consuming 
20W. The GPU equipped in the Amdahl blade is another solution. We can 
accelerate the Hadoop framework and the MapReduce application by offloading 
CPU-heavy operations to GPU. For example, it is worth offloading compression, 
checksum calculation and verification, and data sorting during data shuffling to GPU.

%% file: concl.tex
\section{Conclusion}
\label{sec:concl}

The paper shows that Amdahl blades are much more energy-efficient than
a regular Beowulf cluster, in both data-intensive and
compute-intensive Hadoop applications. However, the Amdahl blades are
not well balanced for Hadoop because HDFS employs many network I/O
operations and the Atom processors are the bottleneck. As we
demonstrated, disk I/O and network I/O are both CPU-intensive. One can
reduce CPU use for disk operations with direct I/O, but the only
option for reducing the overhead of network I/O right now is to reduce
the amount of data transmitted, for example with LZO compression. In
the end, we estimate that an Amdahl blade needs four cores in order to
be balanced.